\documentclass[oneside,english]{amsart}
\usepackage[T1]{fontenc}
\usepackage[latin9]{inputenc}
\usepackage{geometry}
\usepackage{textcomp}
\usepackage{amstext}
\usepackage{amsthm}
\usepackage{amssymb}
\usepackage{esint}

\makeatletter
\numberwithin{equation}{section}
\numberwithin{figure}{section}

\makeatother

\usepackage{babel}
\begin{document}
\title{Phenomenology from Dirac equation\\
with Euclidean-Minkowskian ``gravity phase''}
\date{16 Jan 2023. This preprint has not undergone peer review or any post-submission
improvements or corrections. The Version of Record of this article
is published in Int. J. Theor. Phys., and is available online at https://doi.org/10.1007/s10773-023-05283-2}
\author{Jens K\"oplinger}
\address{105 E Avondale Dr, Greensboro, NC 27403, USA}
\email{jenskoeplinger@gmail.com}
\urladdr{preprints, personal versions at http://jenskoeplinger.com/P}
\begin{abstract}
Over the past decades, many authors advertised models on complexified
spacetime algebras for use in describing gravity. This work aims at
providing phenomenological support to such claims, by introducing
a one-parameter real phase $\alpha$ to the conventional Dirac equation
with $\frac{1}{r}$-type potential. This phase allows to transition
between Euclidean ($\alpha=0,\pm\pi,\pm2\pi,\ldots$) and Minkowskian
($\alpha=\pm\frac{\pi}{2},\pm\frac{3\pi}{2},\ldots$) geometry, as
two distinct cases that one may expect from some complexified spacetime.
The configuration space is modeled on $4\times4$ matrix algebra over
the bicomplex numbers, $\mathbb{C}\oplus\mathbb{C}$. Spin-$\frac{1}{2}$
Coulomb scattering (Rutherford scattering) in Born approximation is
then executed. All calculations are done ``from scratch'', as they
could have been done some 85 years ago. By removing elegance from
field theory that has since become customary, this paper aims at remaining
as generally applicable as possible, for a wide range of candidate
models that contain such a phase $\alpha$ in one way or another.
Results for backscattering and cross section at high energies are
compared with results from General Relativity calculations. Effects
on intergalactic gas distribution and momentum transfer from scattering
high-energy leptons are sketched.\tableofcontents{}
\end{abstract}

\maketitle

\section{Context}

Numerous models have been proposed in mathematical physics over the
past decades, which provide ways that one could interpret as ``complexifying
spacetime'' algebraically. Out of those, several claim to be modelling
effects from gravity this way. For a certainly not complete list see
e.g. \cite{AymazKanzu2021,BaylisKeselica2012,Bronoff2011,CastroPerelman2021,ChanyalSharmaNegi2015,Demir2013ijtp,Demir2013,DemirTanisli2012,DemirTanisliTolan2013,DuffMadore1978,Gu2022,koepl_1_lDiracEqnOnSplitO2006,koepl_2_SignOfGrav2007,koepl_3_HypernumbersRel2007,koepl_5_QuantumOfArea2008,MK2020,MironovMironov2014,MironovMironov2016,MironovMironov2017,Panicaud2011,TanisliKansuDemir2014,Ulrych2005,Ulrych2006,VelezQuinonew2022,Weng2009}.
These works don't necessarily assume locally-Lorentzian spacetime
as foundational. They are therefore distinct from canonical ``spin-$2$''
type gravity, as well as, use of four-dimensional Euclidean spacetime
as mathematical tool in contemporary field theory (keywords ``$ict$'',
``Wick rotation'').

In order to support taking such a stance a priori, this paper provides
a simple phenomenological calculation: In the spirit of retracing
early quantum electrodynamics \cite{Miller1994}, mathematical elegance
is removed as much as possible, leaving just enough structure to be
able to arrive at qualitative and quantitative results. The hope is
that these results could then be reproduced in the respective algebraic
frameworks of a wide range of models, if so desired.

Complexifying a conventional $4\times4$ complex matrix representation
of the Dirac equation can be understood as matrix algebra over the
bicomplex numbers $\mathbb{C}\oplus\mathbb{C}$, which have rich historical
and contemporary interest in physics (for pointers see e.g. \cite{WPbicomplex}).

\section{Introduce a phase $\alpha$ to the Dirac equation}

We start with the conventional Dirac equation and a $\frac{1}{r}$-type
potential, and introduce a one-parameter real phase $\alpha$ that
transitions between Euclidean ($\alpha=0,\pm\pi,\pm2\pi,\ldots$)
and Minkowskian ($\alpha=\pm\frac{\pi}{2},\pm\frac{3\pi}{2},\ldots$)
geometry. This phase is proposed to serve as the minimal assumption
that justifies utility of the calculation in ``complexified spacetime'',
which contains Euclidean and Lorentzian subspaces, respectively.

\subsection{Define phase $\alpha$}

Let $\alpha\in\mathbb{R}$ be the phase that transitions between Euclidean
($\alpha=0,\pm\pi,\pm2\pi,\ldots$) and Minkowskian ($\alpha=\pm\frac{\pi}{2},\pm\frac{3\pi}{2},\ldots$)
geometry. Define two copies of the complex numbers, $\mathbb{C}$
and $\mathbb{C}_{0}$, to basis elements $\left\{ 1,i\right\} $ and
$\left\{ 1,i_{0}\right\} $, respectively. Dedicate $\mathbb{C}_{0}$
to modeling the phase from $\alpha$, and $\mathbb{C}$ for modeling
conventional Dirac equation and operators. Elements from $\mathbb{C}$
and $\mathbb{C}_{0}$ commute, associate, and distribute just like
real coefficients. Given a variable $x$, complex conjugation in $\mathbb{C}$
is written as $\overline{x}$ and complex conjugation in $\mathbb{C}_{0}$
as $\underline{x}$.

With this, a transformation $\phi$ and its conjugate $\underline{\phi}$
are then defined as:
\begin{equation}
\phi\,:=\,e^{i_{0}\alpha},\qquad\underline{\phi}\,:=\,e^{-i_{0}\alpha},\qquad\phi\underline{\phi}=\underline{\phi}\phi=1.
\end{equation}

\subsection{Introduce $\alpha$ to Dirac equation}

\label{subsec:introduceAlphaToMatrices}Using $4\times4$ matrices
over the complexes, $\gamma_{\mu}$, with $\mu=0\ldots3$, linear
derivatives $\partial_{\mu}:=\frac{\partial_{\mu}}{\partial x_{\mu}}$,
a property $\widetilde{m}$ that is invariant under $\partial_{\mu}$
(with $\left.\widetilde{m}\right|_{\mathrm{QED}}\equiv m\in\mathbb{R}$
mass in the classical case), and functions $\psi:\mathbb{R}^{4}\rightarrow\mathbb{C}^{4}$,
the Dirac equation is written as the eigenvalue relation:
\begin{equation}
\sum_{\mu=0}^{3}i\gamma_{\mu}\partial_{\mu}\psi=\widetilde{m}\psi.\label{eq:DiracEqn}
\end{equation}
In contrast to notation convention in physics today, all indices are
now written as lower indices, and summation is written explicitly
(i.e. without implicit Minkowski tensor). Spelling out summations
and metric explicitly avoids confusion going forward, when Minkowski
metric is considered an edge case in a generalized geometry.

Generalized Dirac matrices, using the same symbol $\gamma_{\mu}$,
are now defined as a function of $\alpha$, to model the conventional
Dirac equation in the $\alpha=\pm\frac{\pi}{2},\pm\frac{3\pi}{2},\ldots$
cases and a counterpart on Euclidean metric in the $\alpha=0,\pm\pi,\pm2\pi,\ldots$
case (per \cite{koepl_1_lDiracEqnOnSplitO2006,koepl_2_SignOfGrav2007}):
\begin{equation}
\gamma_{0}:=\left(\begin{array}{cccc}
1 & 0 & 0 & 0\\
0 & 1 & 0 & 0\\
0 & 0 & -1 & 0\\
0 & 0 & 0 & -1
\end{array}\right),\,\gamma_{1}:=\left(\begin{array}{cccc}
0 & 0 & 0 & \phi^{2}\\
0 & 0 & \phi^{2} & 0\\
0 & 1 & 0 & 0\\
1 & 0 & 0 & 0
\end{array}\right),\,\gamma_{2}:=\left(\begin{array}{cccc}
0 & 0 & 0 & -i\phi^{2}\\
0 & 0 & i\phi^{2} & 0\\
0 & -i & 0 & 0\\
i & 0 & 0 & 0
\end{array}\right),\,\gamma_{3}:=\left(\begin{array}{cccc}
0 & 0 & \phi^{2} & 0\\
0 & 0 & 0 & -\phi^{2}\\
1 & 0 & 0 & 0\\
0 & -1 & 0 & 0
\end{array}\right).
\end{equation}
With $\gamma_{\mu}\in\left(\mathbb{C}\times\mathbb{C}_{0}\right)^{4}$
the wave functions $\psi$ are now generally $\psi:\mathbb{R}^{4}\rightarrow\left(\mathbb{C}\times\mathbb{C}_{0}\right)^{4}$.

Using Pauli spinors $\sigma_{j}$ with $j=1,2,3$
\begin{equation}
\sigma_{1}:=\left(\begin{array}{cc}
0 & 1\\
1 & 0
\end{array}\right),\qquad\sigma_{2}:=\left(\begin{array}{cc}
0 & -i\\
i & 0
\end{array}\right),\qquad\sigma_{3}:=\left(\begin{array}{cc}
1 & 0\\
0 & -1
\end{array}\right),
\end{equation}
and identifying the unit matrix with
\begin{eqnarray}
I_{2}\equiv\sigma_{0} & := & \left(\begin{array}{cc}
1 & 0\\
0 & 1
\end{array}\right),
\end{eqnarray}
the generalized $\gamma_{\mu}$ can be written as:
\begin{equation}
\gamma_{0}=\left(\begin{array}{cc}
\sigma_{0} & 0\\
0 & -\sigma_{0}
\end{array}\right)=\left(\begin{array}{cc}
I_{2} & 0\\
0 & -I_{2}
\end{array}\right),\qquad\gamma_{j}=\left(\begin{array}{cc}
0 & \phi^{2}\sigma_{j}\\
\sigma_{j} & 0
\end{array}\right).
\end{equation}

It is left open for now whether this requires the parameter $\widetilde{m}$
to become complex in $\mathbb{C}_{0}$ or not.

\subsection{Properties of the generalized Dirac-$\gamma$}

The Euclidean and Minkowskian edge cases from the referenced papers
(\cite{koepl_1_lDiracEqnOnSplitO2006,koepl_2_SignOfGrav2007}) are
satisfied by inspection, for $\phi^{2}=1$ and $\phi^{2}=-1$, respectively.

Writing $I_{4}$ for the identity $4\times4$ matrix, the generalized
Dirac matrices have the property:
\begin{align}
\frac{1}{2}\left(\gamma_{\mu}\gamma_{\nu}+\gamma_{\nu}\gamma_{\mu}\right) & =I_{4}*\begin{cases}
0 & \textrm{for }\mu\neq\nu,\\
1 & \textrm{for }\mu=\nu=0,\\
\phi^{2} & \textrm{otherwise }\left(\mu=\nu\in\left\{ 1,2,3\right\} \right).
\end{cases}\label{eq:generalizedGammaAnticommutatorRel}
\end{align}
This property reflects the choice of metric.

\subsection{Energy, mass, momentum, relativity}

Understanding the derivatives $i\partial_{\mu}$ as quantum mechanical
operators for energy $E$ and momentum $\vec{p}:=\left(p_{1},p_{2},p_{3}\right)$,
$\left|\vec{p}\right|^{2}:=p_{1}^{2}+p_{2}^{2}+p_{3}^{2}$, multiplying
the Dirac equation (\ref{eq:DiracEqn}) with its conjugate in $\mathbb{C}$
recovers the Minkowskian $\left|E\right|^{2}-\left|\vec{p}\right|^{2}=m_{\mathrm{Mink}}^{2}$
and Euclidean $\left|E\right|^{2}+\left|\vec{p}\right|^{2}=m_{\mathrm{Eucl}}^{2}$
edge cases for $\phi^{2}=1$ and $\phi^{2}=-1$, respectively.

In general, the sum over four momentum $p$,
\begin{equation}
p:=\left(p_{\mu}\right)=\left(E,\vec{p}\right)=\left(E,p_{1},p_{2},p_{3}\right),\qquad\textrm{with }E,\vec{p}\in\mathbb{R}.
\end{equation}
becomes:
\begin{eqnarray}
m_{0}^{2} & := & \sum_{\mu=0}^{3}\sum_{\nu=0}^{3}\gamma_{\mu}\gamma_{\nu}p_{\mu}p_{\nu}=E^{2}+\phi^{2}\left|\vec{p}\right|^{2}.\label{eq:defMAlphaSquare}
\end{eqnarray}
This makes $m_{0}^{2}$ complex in $\mathbb{C}_{0}$ in the general
case. In order to keep the classical mass parameter $m_{\mathbb{R}}\in\mathbb{R}$
real, it has to be redefined as compared to the conventional case.
The simplest way is to take the absolute of $m_{0}^{2}$, which makes
$m_{\mathbb{R}}$ a fourth-order expression in $E$ and $\vec{p}$:
\begin{eqnarray}
\phi^{2} & = & e^{2i_{0}\alpha},\textrm{ therefore}\\
m_{\mathbb{R}}^{4}: & = & \left|m_{0}^{2}\right|^{2}=E^{4}+2E^{2}\left|\vec{p}\right|^{2}\cos\left(2\alpha\right)+\left|\vec{p}\right|^{4},\\
m_{\mathbb{R}} & = & \sqrt{\left|m_{0}^{2}\right|}=\sqrt[4]{E^{4}+2E^{2}\left|\vec{p}\right|^{2}\cos\left(2\alpha\right)+\left|\vec{p}\right|^{4}}.\label{eq:InvariantMassDef}
\end{eqnarray}
There are both physical and mathematical implications to this.

\subsubsection{Physical implications}

From the physical side, it raises the question on what makes physical
lab frames equivalent. Next to energy and momentum of test bodies
in unaccelerated frames of reference, the parameter $\alpha$ now
factors into the equivalence condition as well. Equation \ref{eq:InvariantMassDef}
therefore becomes the new definition of relativity\footnote{A semi-classical approximation allows to reconstruct linearized General
Relativity, per ``proposition 4'' in \cite{koepl_3_HypernumbersRel2007}
(``\emph{NatAliE} equations''). Ignoring the complex-octonion setting
in that paper, this ``proposition 4'' can be interpreted as stand-alone
phenomenology in the large body, non-quantum limit. This requires
clarification of the meaning of $\alpha$, which is subject of this
current research.}.

The relation between speed $\vec{v}$, energy $E$, and momentum $\vec{p}$
remains unchanged for any $\alpha$ by definition:
\begin{eqnarray}
\vec{v} & := & \frac{\vec{p}}{E}.\label{eq:speedMomentumEnergy}
\end{eqnarray}
Conserved properties between equivalent frames of reference are generalized
per equation (\ref{eq:defMAlphaSquare}) and therefore complex in
$\mathbb{C}_{0}$ in the general case. For example, invariant length
elements $\ell_{0}$, volumes $V_{0}$, or time intervals $d\tau$:
\begin{equation}
\ell_{0}:=\ell\frac{E}{m_{0}}=\frac{\ell}{\sqrt{1+\phi^{2}\left|\vec{v}\right|^{2}}},\qquad V_{0}:=V\frac{E}{m_{0}}=\frac{V}{\sqrt{1+\phi^{2}\left|\vec{v}\right|^{2}}},\qquad d\tau:=dt\frac{m_{0}}{E}=dt\sqrt{1+\phi^{2}\left|\vec{v}\right|^{2}}.\label{eq:defInvariantsGeneralized}
\end{equation}

\subsubsection{Mathematical implications}

From the mathematical side, fourth-order expressions are a departure
from the pure (Dirac-)spinor expression of the equation of motion
of a spin-$\frac{1}{2}$ particle. It requires clarification on what
mathematical construct we're looking at exactly - or alternatively,
find a more natural mathematical representation of the phase $\alpha$.
This may ultimately lead to a more natural mathematical description
of relativity, which might not be apparent here due to the focus on
a special case.

For now it is still left open whether the placeholder $\widetilde{m}$
in the generalized Dirac equation is the classical, real-valued mass
parameter $m_{\mathbb{R}}$, or whether it has to become $m_{0}$
which is complex-valued in $\mathbb{C}_{0}$.

\subsection{Green's function in energy-momentum space}

In momentum space, Green's function $G\left(p\right)$ solves
\begin{equation}
\left(\sum_{\mu=0}^{3}\gamma_{\mu}p_{\mu}-\widetilde{m}\right)G\left(p\right)=1
\end{equation}
and is:
\begin{eqnarray}
G\left(p\right) & = & \frac{\sum_{\mu=0}^{3}\gamma_{\mu}p_{\mu}+\widetilde{m}}{E^{2}+\phi^{2}\left|\vec{p}\right|^{2}-\widetilde{m}^{2}}=\frac{\sum_{\mu=0}^{3}\gamma_{\mu}p_{\mu}+\widetilde{m}}{m_{0}^{2}\left(p\right)-\widetilde{m}^{2}}.
\end{eqnarray}

The factor $m_{0}^{2}\left(p\right)\in\mathbb{C}_{0}$ varies not
only in $\alpha$ but also in energy and momentum. If $\widetilde{m}$
would be assumed real, it would mean that there is no pole in $G\left(p\right)$
on the real $p_{\mu}$ parameter space except for the Euclidean ($\phi^{2}=1$,
$\alpha=0,\pm\pi,\pm2\pi,\ldots$) and Minkowskian ($\phi^{2}=-1$,
$\alpha=\pm\frac{\pi}{2},\pm\frac{3\pi}{2},\ldots$) edge cases.

This raises questions, coming from conventional theory where the propagator
of a plane-wave particle would be expected to have a pole at the particle's
invariant mass $m_{\mathbb{R}}$. There it would mark the exact momentum
and energy associated with the infinite plane wave. Invariant mass
$m_{\mathbb{R}}$ is a particle property, just as is $\alpha$. As
such, it seems from Green's function that it is not correct to model
the generalized Dirac equation using the real mass property $\widetilde{m}\overset{?}{\equiv}m_{\mathbb{R}}$,
but instead the complex mass property $\widetilde{m}\overset{?}{\equiv}\sqrt{m_{0}^{2}}$
should be used. Nevertheless, for now the placeholder $\widetilde{m}$
continues to be used until more evidence is gathered.

\section{Rutherford scattering of a spin-$\frac{1}{2}$ particle for general
$\alpha$}

In order to check for plausibility of the overall approach, and to
get a feeling on what challenges may arise when generalizing the calculation,
execute a simple quantum calculation in a special case. Observe how
the phase $\alpha$ plays into that special case, and learn how this
may later need to be handled generally.

Spin-$\frac{1}{2}$ Coulomb scattering (Rutherford scattering) in
Born approximation is a simple and well understood calculation that
can be done from the Dirac equation with minimal prerequisites. The
calculation will correct and clean up the $\alpha=0$ case in \cite{koepl_5_QuantumOfArea2008},
but then also provide the result for general alpha:
\begin{itemize}
\item Fix a representation of the Dirac equation and introduce the phase
$\alpha$ in the simplest possible way.
\item Define some basics (adjoint wave function, conservation of probability,
particle propagator).
\item Use Born approximation to calculate cross section.
\end{itemize}
Interpret the result:
\begin{itemize}
\item Backscattering, cross section for high energies,
\item compare with cross section of \textquotedbl scattering a rock on
a Black Hole\textquotedbl ,
\item estimate effects on intergalactic gas distribution,
\item estimate momentum transfer $\triangle p$ from scattering high-energy
neutrinos on atomic nuclei when traveling through matter.
\end{itemize}

\subsection{Adjoint wave function and conservation of probability}

For a given volume and time interval without sources, probability
must be conserved in the general case. This is a prerequisite for
calculating transition properties using perturbation theory methods,
but is also a physical principle underlying quantum mechanics in general.

Define a matrix $\gamma_{\phi}$ as:
\begin{eqnarray}
\gamma_{\phi} & := & \left(\begin{array}{cccc}
1 & 0 & 0 & 0\\
0 & 1 & 0 & 0\\
0 & 0 & \phi & 0\\
0 & 0 & 0 & \phi
\end{array}\right)=\left(\begin{array}{cc}
\sigma_{0} & 0\\
0 & \phi\sigma_{0}
\end{array}\right),\qquad\gamma_{\phi}^{2}=\gamma_{\phi^{2}},
\end{eqnarray}
with
\begin{eqnarray}
\gamma_{\underline{\phi}} & = & \left(\begin{array}{cccc}
1 & 0 & 0 & 0\\
0 & 1 & 0 & 0\\
0 & 0 & \underline{\phi} & 0\\
0 & 0 & 0 & \underline{\phi}
\end{array}\right)=\left(\begin{array}{cc}
\sigma_{0} & 0\\
0 & \underline{\phi}\sigma_{0}
\end{array}\right),\qquad\gamma_{\underline{\phi}}^{2}=\gamma_{\underline{\phi}^{2}},
\end{eqnarray}
accordingly. Therefore
\begin{eqnarray}
\gamma_{\underline{\phi}}\gamma_{\phi}=\gamma_{\phi}\gamma_{\underline{\phi}}=\gamma_{\underline{\phi}^{2}}\gamma_{\phi^{2}}=\gamma_{\phi^{2}}\gamma_{\underline{\phi}^{2}} & = & I_{4}.
\end{eqnarray}
It relates a $\gamma_{\mu}$ matrix with its $\mathbb{C}$-hermitian
transpose, $\overline{\gamma}_{\mu}^{T}$:
\begin{equation}
\gamma_{\mu}=\gamma_{\underline{\phi}^{2}}\overline{\gamma}_{\mu}^{T}\gamma_{\phi^{2}},\qquad\overline{\gamma}_{\mu}^{T}=\gamma_{\phi^{2}}\gamma_{\mu}\gamma_{\underline{\phi}^{2}}.
\end{equation}

Writing $\psi^{\dagger}$ for adjoint and $\psi^{T}$ for transpose
of $\psi$, all $\mathbb{R}^{4}\rightarrow\left(\mathbb{C}\times\mathbb{C}_{0}\right)^{4}$,
the probability density four-vector $j:\mathbb{R}^{4}\rightarrow\mathbb{R}^{4}$
is defined as:
\begin{eqnarray}
j_{\mu} & := & \psi^{\dagger}\gamma_{\mu}\psi\textrm{ with }\mu=0\ldots3.
\end{eqnarray}
In the absence of sources (charges and fields) probability must be
conserved globally:
\begin{eqnarray}
\sum_{\mu=0}^{3}\partial_{\mu}j_{\mu} & \overset{!}{=} & 0.
\end{eqnarray}
This fixes the adjoint as:
\begin{eqnarray}
\psi^{\dagger} & := & \overline{\psi}^{T}\gamma_{\phi^{2}}.
\end{eqnarray}

\begin{proof}
Build the $\mathbb{C}$-hermitian transpose of the Dirac equation,
$\sum_{\mu=0}^{3}i\gamma_{\mu}\partial_{\mu}\psi=\widetilde{m}\psi$.
Conjugation is done in $\mathbb{C}$ since the quantum mechanical
probability amplitude is modeled in this subalgebra. The differentials
$\partial_{\mu}$ are understood as acting on the wave function $\overline{\psi}^{T}$,
i.e., to the left in this case:
\begin{eqnarray}
\overline{\psi}^{T}\sum_{\mu=0}^{3}\left(-i\overline{\gamma}_{\mu}^{T}\partial_{\mu}\right) & = & \overline{\psi}^{T}\widetilde{m},\\
\overline{\psi}^{T}\sum_{\mu=0}^{3}\left(-i\gamma_{\phi^{2}}\gamma_{\mu}\gamma_{\underline{\phi}^{2}}\partial_{\mu}\right) & = & \overline{\psi}^{T}\widetilde{m}.
\end{eqnarray}
Pull out $\gamma_{\phi^{2}}$ to the left, identify $\widetilde{m}=\gamma_{\phi^{2}}\widetilde{m}\gamma_{\underline{\phi}^{2}}$,
then multiply with $\gamma_{\phi^{2}}$ from the right and identify
the adjoint $\psi^{\dagger}$:
\begin{eqnarray}
\overline{\psi}^{T}\gamma_{\phi^{2}}\sum_{\mu=0}^{3}\left(-i\gamma_{\mu}\gamma_{\underline{\phi}^{2}}\partial_{\mu}\right) & = & \overline{\psi}^{T}\gamma_{\phi^{2}}\widetilde{m}\gamma_{\underline{\phi}^{2}},\\
\overline{\psi}^{T}\gamma_{\phi^{2}}\sum_{\mu=0}^{3}\left(-i\gamma_{\mu}\partial_{\mu}\right) & = & \overline{\psi}^{T}\gamma_{\phi^{2}}\widetilde{m},\\
\psi^{\dagger}\sum_{\mu=0}^{3}\left(-i\gamma_{\mu}\partial_{\mu}\right) & = & \psi^{\dagger}\widetilde{m},\\
\sum_{\mu=0}^{3}\left(\left(\partial_{\mu}\psi^{\dagger}\right)\gamma_{\mu}\right) & = & i\widetilde{m}\psi^{\dagger}.
\end{eqnarray}
The last line only reordered the terms. With this:
\begin{equation}
\sum_{\mu=0}^{3}\partial_{\mu}j_{\mu}=\sum_{\mu=0}^{3}\partial_{\mu}\left(\psi^{\dagger}\gamma_{\mu}\psi\right)=\sum_{\mu=0}^{3}\left(\left(\partial_{\mu}\psi^{\dagger}\gamma_{\mu}\right)\psi+\psi^{\dagger}\left(\gamma_{\mu}\partial_{\mu}\psi\right)\right)=i\widetilde{m}\psi^{\dagger}\psi-i\widetilde{m}\psi^{\dagger}\psi=0.
\end{equation}
Probability density is conserved.
\end{proof}
Note that this proof holds regardless of whether $\widetilde{m}$
is real-valued or complex-valued in $\mathbb{C}_{0}$. The restriction
on $\widetilde{m}$ is that it must not be complex-valued in $\mathbb{C}$,
since complex conjugation in that space is used for the proof.

The classical adjoint $\left.\psi^{\dagger}\right|_{\mathrm{QED}}$
is recovered as expected for $\phi^{2}=-1$ as
\begin{eqnarray}
\left.\psi^{\dagger}\right|_{\mathrm{QED}} & = & \overline{\psi}^{T}\left.\gamma_{0}\right|_{\mathrm{QED}}=\overline{\psi}^{T}\left(\begin{array}{cc}
\sigma_{0} & 0\\
0 & -\sigma_{0}
\end{array}\right).
\end{eqnarray}

\subsection{Solutions of the free equation of motion}

The free Dirac equation, $\sum_{\mu=0}^{3}i\gamma_{\mu}\partial_{\mu}\psi=\widetilde{m}\psi$
(\ref{eq:DiracEqn}) is explicitly:
\begin{eqnarray}
i\left[\sum_{\mu=0}^{3}\gamma_{\mu}\partial_{\mu}\right]\psi & = & \widetilde{m}\psi,\\
i\left[\left(\begin{array}{cc}
\sigma_{0} & 0\\
0 & -\sigma_{0}
\end{array}\right)\partial_{0}+\sum_{j=1}^{3}\left(\begin{array}{cc}
0 & \phi^{2}\sigma_{j}\\
\sigma_{j} & 0
\end{array}\right)\partial_{j}\right]\psi & = & \widetilde{m}\psi,\\
i\left[\left(\begin{array}{cccc}
1 & 0 & 0 & 0\\
0 & 1 & 0 & 0\\
0 & 0 & -1 & 0\\
0 & 0 & 0 & -1
\end{array}\right)\partial_{0}+\left(\begin{array}{cccc}
0 & 0 & 0 & \phi^{2}\\
0 & 0 & \phi^{2} & 0\\
0 & 1 & 0 & 0\\
1 & 0 & 0 & 0
\end{array}\right)\partial_{1}+\right.\\
\left.\left(\begin{array}{cccc}
0 & 0 & 0 & -i\phi^{2}\\
0 & 0 & i\phi^{2} & 0\\
0 & -i & 0 & 0\\
i & 0 & 0 & 0
\end{array}\right)\partial_{2}+\left(\begin{array}{cccc}
0 & 0 & \phi^{2} & 0\\
0 & 0 & 0 & -\phi^{2}\\
1 & 0 & 0 & 0\\
0 & -1 & 0 & 0
\end{array}\right)\partial_{3}\right]\psi & = & \widetilde{m}\psi,\nonumber \\
\left(\begin{array}{cccc}
i\partial_{0} & 0 & \phi^{2}\left(i\partial_{3}\right) & \phi^{2}\left(i\partial_{1}+\partial_{2}\right)\\
0 & i\partial_{0} & \phi^{2}\left(i\partial_{1}-\partial_{2}\right) & \phi^{2}\left(-i\partial_{3}\right)\\
i\partial_{3} & i\partial_{1}+\partial_{2} & -i\partial_{0} & 0\\
i\partial_{1}-\partial_{2} & -i\partial_{3} & 0 & -i\partial_{0}
\end{array}\right)\psi & = & \widetilde{m}\psi.\label{eq:generalizedDiracExplicitelyMatrixForm}
\end{eqnarray}

Finding eingenfunctions and eigenvalues requires an energy-momentum-mass
relation, which is $\left|E\right|^{2}-\left|\vec{p}\right|^{2}=m_{\mathrm{Mink}}^{2}$
in the classical case, and $\left|E\right|^{2}+\left|\vec{p}\right|^{2}=m_{\mathrm{Eucl}}^{2}$
in the 4D Euclidean case (\cite{koepl_2_SignOfGrav2007}). This obviously
has to be generalized now due to the phase $\alpha$ contained in
$\phi^{2}$. Equation (\ref{eq:defMAlphaSquare}) has the consistent
generalization,
\begin{eqnarray}
m_{0}^{2} & = & E^{2}+\phi^{2}\left|\vec{p}\right|^{2}.
\end{eqnarray}

This is not consistent any more with a real-valued $\left.\widetilde{m}\right|_{\mathrm{QED}}\equiv m_{\mathbb{R}}\in\mathbb{R}$
as in the classical case. Eigenfunctions of this linear differential
equation - to be found - must contain an exponential function part
$\sim\exp\left(f\left(p\right)\right)$, as well as a vector part.
The vector part will contain terms of $\vec{p}$ and $E$ that multiply
with additional terms $\vec{p}$ and $E$ from the differential on
the exponential function part. In order for $\widetilde{m}$ to be
real-valued, all factors $\phi^{2}$ would have to cancel out directly
from these differentials. However, since all momentum differentials
in (\ref{eq:generalizedDiracExplicitelyMatrixForm}) appear both with
and without factors $\phi^{2}$, this is impossible in principle.
Therefore, $\widetilde{m}$ must be complex-valued in $\mathbb{C}_{0}$
and the placeholder $\widetilde{m}$ is determined to be $m_{0}$
(and not $m_{\mathbb{R}}$) going forward:
\begin{eqnarray}
\widetilde{m} & := & m_{0},\\
i\left[\sum_{\mu=0}^{3}\gamma_{\mu}\partial_{\mu}\right]\psi & = & m_{0}\psi.
\end{eqnarray}

There is an underlying choice that is made to arrive at this identification:
The generalized Dirac equation is to remain as similar as possible
to the classical case, with as little as needed modification in the
formulation as possible.

With this, eigenfunctions $\widetilde{\Psi}_{1/2}^{\pm}$ can be found:
\begin{align}
\widetilde{\Psi}_{1}^{+} & :=\exp i\left(\vec{p}\vec{x}-Et\right)\left(\begin{array}{c}
1\\
0\\
-p_{3}/\left(m_{0}+E\right)\\
\left(-p_{1}-ip_{2}\right)/\left(m_{0}+E\right)
\end{array}\right),\qquad\widetilde{\Psi}_{1}^{-}:=\exp i\left(\vec{p}\vec{x}-Et\right)\left(\begin{array}{c}
0\\
1\\
\left(-p_{1}+ip_{2}\right)/\left(m_{0}+E\right)\\
p_{3}/\left(m_{0}+E\right)
\end{array}\right),\label{eq:genDirPsi1PlusMinus}\\
\widetilde{\Psi}_{2}^{+} & :=\exp i\left(\vec{p}\vec{x}+Et\right)\left(\begin{array}{c}
-\phi^{2}p_{3}/\left(m_{0}+E\right)\\
\phi^{2}\left(-p_{1}-ip_{2}\right)/\left(m_{0}+E\right)\\
1\\
0
\end{array}\right),\qquad\widetilde{\Psi}_{2}^{-}:=\exp i\left(\vec{p}\vec{x}+Et\right)\left(\begin{array}{c}
\phi^{2}\left(-p_{1}+ip_{2}\right)/\left(m_{0}+E\right)\\
\phi^{2}p_{3}/\left(m_{0}+E\right)\\
0\\
1
\end{array}\right).\label{eq:genDirPsi2PlusMinus}
\end{align}

In comparison to the classical solutions ($\phi^{2}=-1$) the $\widetilde{\Psi}_{1/2}^{\pm}$
can be identified as (anti)particle ($1/2$) plane waves with spin
up ($+$) or down ($-$). Here in the general case, $m_{0}$ varies
in $\mathbb{C}_{0}$ for all solutions, and there is a phase $\phi^{2}$
in the vector parts of the eigenfunctions between particles and antiparticles.

Using a spin vector $\chi^{\pm}$ with
\begin{equation}
\chi^{+}:=\left(\begin{array}{c}
1\\
0
\end{array}\right),\qquad\qquad\chi^{-}:=\left(\begin{array}{c}
0\\
1
\end{array}\right),
\end{equation}
the eigenfunctions can be expressed using Pauli spinors $\sigma_{\mu}$
and $\vec{\sigma}:=\left(\sigma_{1},\sigma_{2},\sigma_{3}\right)$
as:
\begin{eqnarray}
\widetilde{\Psi}_{1}^{\pm} & := & \exp i\left(\vec{p}\vec{x}-Et\right)\left(\begin{array}{c}
\sigma_{0}\\
-\vec{\sigma}\vec{p}/\left(m_{0}+E\right)
\end{array}\right)\chi^{\pm},\\
\widetilde{\Psi}_{2}^{\pm} & := & \exp i\left(\vec{p}\vec{x}+Et\right)\left(\begin{array}{c}
-\phi^{2}\vec{\sigma}\vec{p}/\left(m_{0}+E\right)\\
\sigma_{0}
\end{array}\right)\chi^{\pm}.
\end{eqnarray}

\subsection{Normalizing the plane-wave eigenfunctions}

Norms of numbers $a\in\mathbb{C}\times\mathbb{C}_{0}$ are defined
in $\mathbb{C}$, $\mathbb{C}_{0}$, and $\mathbb{C}\times\mathbb{C}_{0}$
as:
\begin{align}
\left|a\right|^{2} & :=a\overline{a}\in\mathbb{C}_{0},\qquad\qquad\left|a\right|_{0}^{2}:=a\underline{a}\in\mathbb{C},\qquad\qquad\left\Vert a\right\Vert ^{4}:=a\overline{a}\underline{a\overline{a}}\in\mathbb{R}.
\end{align}
The term ``norm'' is used loosely, in the sense that the composition
property is conserved for any $a,b\in\mathbb{C}\times\mathbb{C}_{0}$:
\begin{equation}
\left|ab\right|^{2}=\left|a\right|^{2}\left|b\right|^{2},\qquad\qquad\left|ab\right|_{0}^{2}=\left|a\right|_{0}^{2}\left|b\right|_{0}^{2},\qquad\qquad\left\Vert ab\right\Vert ^{4}=\left\Vert a\right\Vert ^{4}\left\Vert b\right\Vert ^{4}.
\end{equation}
Norms are, however, not positive definite or point separating. Note
that when taking the square (or fourth) root of these expressions
it has to be made clear which space the result is to be in ($\mathbb{C}$,
$\mathbb{C}_{0}$, or $\mathbb{C}\times\mathbb{C}_{0}$).

Since the generalized Dirac equation uses an eigenvalue $m_{0}\in\mathbb{C}_{0}$,
the calculation of the scattering cross section will be executed in
the $\mathbb{C}$ subalgebra only, i.e., in the same manner as in
the classical case. This is possible since all dynamic variables act
in this subalgebra, and the only variable in $\mathbb{C}_{0}$ is
$\alpha$ itself which is constant in space and time (and therewith
in $p$). Only at the very end, when asking for probabilities, will
the $\mathbb{C}_{0}$ norm be taken, to obtain a real value.

The $\widetilde{\Psi}_{1/2}^{\pm}$ above are obtained by fixing components
$\left(1,0\right)$ and $\left(0,1\right)$ in their vector parts.
They are not yet normed to conserve probability in field-free space
(and time). The normed eigenfunctions $\hat{\Psi}_{1/2}^{\pm}$ satisfy
\begin{eqnarray}
\left(\hat{\Psi}_{1/2}^{\pm}\right)^{\dagger}\hat{\Psi}_{1/2}^{\pm} & \overset{!}{=} & 1.
\end{eqnarray}
They differ from the $\widetilde{\Psi}_{1/2}^{\pm}$ only by a constant
factor $N_{1/2}\in\mathbb{C}_{0}$,
\begin{eqnarray}
\hat{\Psi}_{1/2}^{\pm} & := & N_{1/2}\widetilde{\Psi}_{1/2}^{\pm}.
\end{eqnarray}

For $\widetilde{\Psi}_{1}^{+}$there is:
\begin{eqnarray}
\left(\hat{\Psi}_{1}^{+}\right)^{\dagger}\hat{\Psi}_{1}^{+} & = & \overline{\left(\hat{\Psi}_{1}^{+}\right)}^{T}\gamma_{\phi^{2}}\hat{\Psi}_{1}^{+}=\overline{N_{1}}\overline{\left(\widetilde{\Psi}_{1}^{+}\right)}^{T}\gamma_{\phi^{2}}N_{1}\widetilde{\Psi}_{1}^{+}\\
 & = & \left|N_{1}\right|^{2}\left(1,0,\frac{-p_{3}}{m_{0}+E},\frac{-p_{1}+ip_{2}}{m_{0}+E}\right)\left(\begin{array}{cccc}
1 & 0 & 0 & 0\\
0 & 1 & 0 & 0\\
0 & 0 & \phi^{2} & 0\\
0 & 0 & 0 & \phi^{2}
\end{array}\right)\left(\begin{array}{c}
1\\
0\\
-p_{3}/\left(m_{0}+E\right)\\
\left(-p_{1}-ip_{2}\right)/\left(m_{0}+E\right)
\end{array}\right)\\
 & = & \left|N_{1}\right|^{2}\left(1+\frac{\phi^{2}p_{3}^{2}}{\left(m_{0}+E\right)^{2}}+\frac{\phi^{2}\left(p_{1}^{2}+p_{2}^{2}\right)}{\left(m_{0}+E\right)^{2}}\right)\\
 & = & \left|N_{1}\right|^{2}\left(1+\frac{\phi^{2}\left|\vec{p}\right|^{2}}{\left(m_{0}+E\right)^{2}}\right).
\end{eqnarray}
Using $m_{0}^{2}=E^{2}+\phi^{2}\left|\vec{p}\right|^{2}$ this becomes:
\begin{eqnarray}
\left(\hat{\Psi}_{1}^{+}\right)^{\dagger}\hat{\Psi}_{1}^{+} & = & \left|N_{1}\right|^{2}\frac{\left(m_{0}+E\right)^{2}+\phi^{2}\left|\vec{p}\right|^{2}}{\left(m_{0}+E\right)^{2}}\,=\,\left|N_{1}\right|^{2}\frac{m_{0}^{2}+2m_{0}E+E^{2}+\phi^{2}\left|\vec{p}\right|^{2}}{\left(m_{0}+E\right)^{2}}\\
 & = & \left|N_{1}\right|^{2}\frac{m_{0}^{2}+2m_{0}E+m_{0}^{2}}{\left(m_{0}+E\right)^{2}}\,=\,\left|N_{1}\right|^{2}\frac{2m_{0}\left(m_{0}+E\right)}{\left(m_{0}+E\right)^{2}}\\
 & = & \left|N_{1}\right|^{2}\frac{2m_{0}}{m_{0}+E}.
\end{eqnarray}
The normalizing factor $N_{1}\in\mathbb{C}_{0}$ therefore differs
from the classical case only by generalizing real mass $m_{\mathbb{R}}$
to $m_{0}$:
\begin{equation}
\left|N_{1}\right|^{2}=\frac{m_{0}+E}{2m_{0}}\in\mathbb{C}_{0},\qquad\qquad N_{1}=\sqrt{\frac{m_{0}+E}{2m_{0}}}.
\end{equation}
The square root is to be taken in $\mathbb{C}_{0}$. The identical
calculation can be done for the $\widetilde{\Psi}_{1}^{-}$.

For the $\widetilde{\Psi}_{2}^{+}$ we have:
\begin{eqnarray}
\left(\hat{\Psi}_{2}^{+}\right)^{\dagger}\hat{\Psi}_{2}^{+} & = & \left|N_{2}\right|^{2}\left(\frac{-\phi^{2}p_{3}}{m_{0}+E},\frac{-\phi^{2}p_{1}+i\phi^{2}p_{2}}{m_{0}+E},1,0\right)\left(\begin{array}{cccc}
1 & 0 & 0 & 0\\
0 & 1 & 0 & 0\\
0 & 0 & \phi^{2} & 0\\
0 & 0 & 0 & \phi^{2}
\end{array}\right)\left(\begin{array}{c}
-\phi^{2}p_{3}/\left(m_{0}+E\right)\\
\phi^{2}\left(-p_{1}-ip_{2}\right)/\left(m_{0}+E\right)\\
1\\
0
\end{array}\right)\\
 & = & \phi^{2}\left|N_{2}\right|^{2}\left(\frac{\phi^{2}p_{3}^{2}}{\left(m_{0}+E\right)^{2}}+\frac{\phi^{2}\left(p_{1}^{2}+p_{2}^{2}\right)}{\left(m_{0}+E\right)^{2}}+1\right)\\
 & = & \phi^{2}\left|N_{2}\right|^{2}\frac{\phi^{2}\left|\vec{p}\right|^{2}+\left(m_{0}+E\right)^{2}}{\left(m_{0}+E\right)^{2}}\\
 & = & \phi^{2}\left|N_{2}\right|^{2}\frac{2m_{0}}{m_{0}+E}.
\end{eqnarray}
Since $\phi\underline{\phi}=1$ this yields:
\begin{equation}
\left|N_{2}\right|^{2}=\underline{\phi}^{2}\frac{m_{0}+E}{2m_{0}}\in\mathbb{C}_{0},\qquad\qquad N_{2}=\underline{\phi}\sqrt{\frac{m_{0}+E}{2m_{0}}}.
\end{equation}

The constant $N_{2}$ is rotated by a phase $\underline{\phi}$ as
compared to $N_{1}$,r
\begin{eqnarray}
N_{2} & = & \underline{\phi}N_{1}.
\end{eqnarray}
This will be of no consequence in this paper since particles and antiparticles
won't change into one another during elastic scattering. For future
work that investigates interactions and transitions involving both
particles and antiparticles, e.g. fermion pair production and annihilation,
this phase may affect the measurement prediction.

You could of course multiply $\underline{\phi}^{2}$ into the eigenfunctions
$\widetilde{\Psi}_{2}^{\pm}$ themselves:
\begin{eqnarray}
\widetilde{\Psi}_{1}^{\pm} & = & \exp i\left(\vec{p}\vec{x}-Et\right)\left(\begin{array}{c}
\sigma_{0}\\
-\vec{\sigma}\vec{p}/\left(m_{0}+E\right)
\end{array}\right)\chi^{\pm},\\
\widetilde{\Psi}_{2}^{\pm}\underline{\phi}^{2} & = & \exp i\left(\vec{p}\vec{x}+Et\right)\left(\begin{array}{c}
-\vec{\sigma}\vec{p}/\left(m_{0}+E\right)\\
\underline{\phi}^{2}\sigma_{0}
\end{array}\right)\chi^{\pm}.
\end{eqnarray}
While this would make the momentum parts $-\vec{\sigma}\vec{p}/\left(m_{0}+E\right)$
symmetric between the $\widetilde{\Psi}_{1}^{\pm}$ and $\widetilde{\Psi}_{2}^{\pm}\underline{\phi}^{2}$,
the $\sigma_{0}$ parts would become asymmetric. Since there's no
real gain for the purpose of this paper, the definitions are left
unchanged.

The normalized component of the eigenfunctions that depends on the
particle's spin and type, $u_{1/2}^{\pm}$, is defined as:
\begin{eqnarray}
u_{1}^{\pm} & := & N_{1}\left(\begin{array}{c}
\sigma_{0}\\
-\vec{\sigma}\vec{p}/\left(m_{0}+E\right)
\end{array}\right)\chi^{\pm}=\sqrt{\frac{m_{0}+E}{2m_{0}}}\left(\begin{array}{c}
\sigma_{0}\\
-\vec{\sigma}\vec{p}/\left(m_{0}+E\right)
\end{array}\right)\chi^{\pm},\label{eq:defUParticle}\\
 &  & u_{1}^{+}=\sqrt{\frac{m_{0}+E}{2m_{0}}}\left(\begin{array}{c}
1\\
0\\
-p_{3}/\left(m_{0}+E\right)\\
\left(-p_{1}-ip_{2}\right)/\left(m_{0}+E\right)
\end{array}\right),\qquad u_{1}^{-}=\sqrt{\frac{m_{0}+E}{2m_{0}}}\left(\begin{array}{c}
0\\
1\\
\left(-p_{1}+ip_{2}\right)/\left(m_{0}+E\right)\\
p_{3}/\left(m_{0}+E\right)
\end{array}\right),\nonumber \\
u_{2}^{\pm} & := & N_{2}\left(\begin{array}{c}
-\phi^{2}\vec{\sigma}\vec{p}/\left(m_{0}+E\right)\\
\sigma_{0}
\end{array}\right)\chi^{\pm}=\sqrt{\frac{m_{0}+E}{2m_{0}}}\left(\begin{array}{c}
-\phi\vec{\sigma}\vec{p}/\left(m_{0}+E\right)\\
\underline{\phi}\sigma_{0}
\end{array}\right)\chi^{\pm},\label{eq:defUAntiparticle}\\
 &  & u_{2}^{+}=\sqrt{\frac{m_{0}+E}{2m_{0}}}\left(\begin{array}{c}
-\phi p_{3}/\left(m_{0}+E\right)\\
\phi\left(-p_{1}-ip_{2}\right)/\left(m_{0}+E\right)\\
\underline{\phi}\\
0
\end{array}\right),\qquad u_{2}^{-}=\sqrt{\frac{m_{0}+E}{2m_{0}}}\left(\begin{array}{c}
\phi\left(-p_{1}+ip_{2}\right)/\left(m_{0}+E\right)\\
\phi p_{3}/\left(m_{0}+E\right)\\
0\\
\underline{\phi}
\end{array}\right).\nonumber 
\end{eqnarray}
This simply leaves out the oscillating plane-wave component of the
eigenfunctions:
\begin{eqnarray}
\hat{\Psi}_{k}^{\pm} & = & u_{k}^{\pm}\exp i\left(\vec{p}\vec{x}+\left(-1\right)^{k}Et\right)\qquad\textrm{with }k=1,2.
\end{eqnarray}

\subsection{Normalizing to invariant volume}

Following textbook calculation of the elastic (Rutherford) scattering
cross section on a fixed target in Born approximation, the wave functions
$\Psi$ of the incoming ($\mathrm{ini}$) and outgoing ($\mathrm{fin}$)
particle will be normalized to a small invariant volume $V_{0}$ in
the reference frame of the target. Per equation (\ref{eq:defInvariantsGeneralized})
the invariant volume $V_{0}$ is
\begin{eqnarray}
V_{0} & = & \frac{VE}{m_{0}}=\frac{V}{\sqrt{1+\phi^{2}\left|\vec{v}\right|^{2}}},
\end{eqnarray}
and the such normalized $\Psi_{k}^{\pm}$ are:
\begin{eqnarray}
\Psi_{k}^{\pm} & := & \sqrt{\frac{m_{0}}{VE}}\hat{\Psi}_{k}^{\pm}\\
 & = & \sqrt{\frac{m_{0}}{VE}}u_{k}^{\pm}\exp i\left(\vec{p}\vec{x}+\left(-1\right)^{k}Et\right).
\end{eqnarray}
This satisfies
\begin{eqnarray}
\left(\Psi_{k}^{\pm}\right)^{\dagger}\Psi_{k}^{\pm} & = & \frac{1}{V_{0}}.
\end{eqnarray}

\subsection{Coulomb-type field of a point charge}

In the classical $\phi^{2}=-1$ case, the electromagnetic field is
introduced by requiring invariance of the Dirac equation with field
under a simple $\exp\left(iq\chi\right)$ phase. There, $q$ is the
electric charge of the particle under the influence of the field,
and space-time derivatives of $\chi$ are identified as the electromagnetic
potentials $A_{\mu}$ acting on the particle proportionally to $q$
with inertia $m_{\mathbb{R}}$.

This is now generalized analogously, writing placeholder symbols $\widetilde{q}$
and $\widetilde{\chi}$ for now until it is clarified exactly which
space they're in:
\begin{align}
\widetilde{A}_{\mu} & :=\frac{\partial\widetilde{\chi}}{\partial x_{\mu}},\qquad\psi^{\prime}:=e^{i\widetilde{q}\widetilde{\chi}}\psi,\\
\sum_{\mu=0}^{3}\gamma_{\mu}\left(i\partial_{\mu}\right)\psi^{\prime}=m_{0}\psi^{\prime} & \qquad\longrightarrow\qquad e^{i\widetilde{q}\widetilde{\chi}}\sum_{\mu=0}^{3}\gamma_{\mu}\left(i\partial_{\mu}-\widetilde{q}\widetilde{A}_{\mu}\right)\psi=m_{0}\psi^{\prime}.
\end{align}
In the general case there is
\begin{align}
\widetilde{q}\widetilde{A}_{\mu} & \in\mathbb{C}_{0},\\
\sum_{\mu=0}^{3}\gamma_{\mu}\left(i\partial_{\mu}+\widetilde{q}\widetilde{A}_{\mu}\right)\psi & =m_{0}\psi.\label{eq:genDiracWithU1gauge}
\end{align}
Equation (\ref{eq:genDiracWithU1gauge}) is the generalized Dirac
equation, invariant under $\mathrm{U}\left(1\right)$ gauge (in $\mathbb{C}$).

This raises the question where exactly the phase in $\mathbb{C}_{0}$
originates from: Is one of the $\left\{ \widetilde{q},\widetilde{A}_{\mu}\right\} $
real or are both complex in $\mathbb{C}_{0}$? By analogy, the purely
electromagnetic and gravitational edge cases have $\left\{ m_{\mathrm{Eucl}},m_{\mathrm{Mink}}\right\} \in\mathbb{R}$,
and we would expect in the gravitational case for the charge to become
its mass, $\widetilde{q}_{\mathrm{Eucl}}\equiv m_{\mathrm{Eucl}}$.
In the electromagnetic case we expect real charges as well, making
$\left\{ \widetilde{q}_{\mathrm{Eucl}},\widetilde{q}_{\mathrm{Mink}}\right\} \in\mathbb{R}$.
Both edge cases appear, however, unphysical: Purely electromagnetic
interaction would assume a particle that is charged, however, doesn't
interact gravitationally; and purely gravitational interaction would
assume a particle or field that has no kinetic component bound to
Minkowskian spacetime. Addressing these concerns is left for later.

Without needing to make speculations on the dynamics behind the $\widetilde{A}_{\mu}$,
a static Coulomb potential of a charge $\widetilde{Q}$ is now modeled
as:
\begin{align}
 & \widetilde{A}_{0}:=-\frac{\widetilde{Q}}{\left|\vec{x}\right|},\qquad\widetilde{A}_{j}=0\textrm{ otherwise},\\
\Longrightarrow & \qquad\sum_{\mu=0}^{3}\gamma_{\mu}\widetilde{q}\widetilde{A}_{\mu}=-\gamma_{0}\frac{\widetilde{q}\widetilde{Q}}{\left|\vec{x}\right|}.
\end{align}
The generalized Dirac equation for a spin-1/2 particle in a static
Coulomb potential then is:
\begin{equation}
\left(\sum_{\mu=0}^{3}i\gamma_{\mu}\partial_{\mu}-\gamma_{0}\frac{\widetilde{q}\widetilde{Q}}{\left|\vec{x}\right|}\right)\psi=m_{0}\psi^{\prime}.
\end{equation}

\subsection{Lowest-order transition matrix element in Born approximation}

For scattering on the fixed point target, initial (incoming) and final
(outgoing) wave functions $\Psi_{k,\mathrm{ini}}^{\pm}$ and $\Psi_{k,\mathrm{fin}}^{\pm}$
of the spin-1/2 particle are assumed plane waves. Only the lowest-order
transition matrix element $S_{\mathrm{fi}}$ is calculated. Using
this customary approximation, results can then be compared qualitatively
with conventional QED results. Some quantitative estimates will also
be possible.
\begin{align}
\Psi_{k,\mathrm{ini}}^{\pm} & :=\sqrt{\frac{m_{0}}{VE}}u_{k,\mathrm{ini}}^{\pm}\exp i\left(\vec{p}\vec{x}+\left(-1\right)^{k}Et\right),\qquad\Psi_{k,\mathrm{fin}}^{\pm}:=\sqrt{\frac{m_{0}}{VE}}u_{k,\mathrm{fin}}^{\pm}\exp i\left(\vec{p}\vec{x}+\left(-1\right)^{k}Et\right),
\end{align}
For now omitting the annotations $\pm$ and $k$ for readability:
\begin{align}
S_{\mathrm{fi}} & =\left\langle \Psi_{\mathrm{fin}}\right|S\left|\Psi_{\mathrm{ini}}\right\rangle =i\int d^{4}x\Psi_{\mathrm{fin}}^{\dagger}\left(\widetilde{q}\sum_{\mu=0}^{3}\gamma_{\mu}A_{\mu}\right)\Psi_{\mathrm{ini}}\\
 & =i\int d^{4}x\Psi_{\mathrm{fin}}^{\dagger}\left(-\widetilde{q}\gamma_{0}\frac{\widetilde{Q}}{\left|\vec{x}\right|}\right)\Psi_{\mathrm{ini}}\\
 & =-i\widetilde{q}\widetilde{Q}\int d^{4}x\left(\sqrt{\frac{m_{0}}{VE_{\mathrm{fin}}}}u_{\mathrm{fin}}\exp i\left(\vec{p}_{\mathrm{fin}}\vec{x}+\left(-1\right)^{k}E_{\mathrm{fin}}t\right)\right)^{\dagger}\left(\gamma_{0}\frac{1}{\left|\vec{x}\right|}\right)\left(\sqrt{\frac{m_{0}}{VE_{\mathrm{ini}}}}u_{\mathrm{ini}}\exp i\left(\vec{p}_{\mathrm{ini}}\vec{x}+\left(-1\right)^{k}E_{\mathrm{ini}}t\right)\right)\\
 & =-i\frac{\widetilde{q}\widetilde{Q}m_{0}}{V\sqrt{E_{\mathrm{fin}}E_{\mathrm{ini}}}}u_{\mathrm{fin}}^{\dagger}\gamma_{0}u_{\mathrm{ini}}\int d^{4}x\frac{\exp\left[i\sum_{\nu=0}^{3}\left(p_{\nu,\mathrm{fin}}-p_{\nu,\mathrm{ini}}\right)\right]}{\left|\vec{x}\right|}.
\end{align}

As compared to the classical case, the differences are in the constants
$\widetilde{q}$, $\widetilde{Q}$, and $m_{0}$ (all valued in $\mathbb{C}_{0}$),
as well as the $u_{k}^{\pm}$ and their adjoint $\left(u_{k}^{\pm}\right)^{\dagger}:=\overline{u_{k}^{\pm}}^{T}\gamma_{\phi^{2}}$
which contain constant terms in $\mathbb{C}_{0}$ as well. Here, ``constant''
means independent of dynamic variables $\left(\vec{x},t\right)$ and
properties $\left(\vec{p},E\right)$. 

When calculating the transition probability $dW$ for a single particle
into a particular state $dN$ (here, into a volume $V$ and momentum
interval $d^{3}p_{\mathrm{fin}}$),
\begin{equation}
dW=\left\Vert S_{\mathrm{fi}}\right\Vert ^{2}dN,
\end{equation}
the norm in $\mathbb{C}\times\mathbb{C}_{0}$ with $\left\Vert a\right\Vert ^{4}:=a\overline{a}\underline{a}\overline{\underline{a}}$
will be used since it is guaranteed real-valued. Other than this intuitive
generalization, standard methods can be followed for calculating the
spin-independent part $d\widetilde{\sigma}_{\mathrm{class}}/d\Omega$
of the cross section $d\widetilde{\sigma}$ into an angle element
$d\Omega$:
\begin{align}
\frac{d\widetilde{\sigma}}{d\Omega} & =\frac{d\widetilde{\sigma}_{\mathrm{class}}}{d\Omega}\frac{d\widetilde{\sigma}_{\mathrm{spin}}}{d\Omega},\label{eq:crossSectionClassVsSpinContrib}\\
\frac{d\widetilde{\sigma}_{\mathrm{class}}}{d\Omega} & =\frac{\left|\widetilde{q}\widetilde{Q}m_{0}\right|^{2}}{4\left|\vec{p}\right|^{4}\sin^{4}\frac{\theta}{2}}.
\end{align}
In order to obtain the real-valued $d\sigma$ at the end, we'll simply
take its absolute again:
\begin{align}
\frac{d\sigma}{d\Omega} & =\left\Vert \frac{d\widetilde{\sigma}}{d\Omega}\right\Vert \equiv\left|\frac{d\widetilde{\sigma}}{d\Omega}\right|_{0}.
\end{align}

\subsection{Spin contribution}

The spin contribution to the cross section (\ref{eq:crossSectionClassVsSpinContrib})
is
\begin{equation}
\frac{d\widetilde{\sigma}_{\mathrm{spin}}}{d\Omega}=\sum_{k,\pm}\left|u_{\mathrm{fin}}^{\dagger}\gamma_{0}u_{\mathrm{ini}}\right|^{2}.
\end{equation}
The symbols $k$ and $\pm$ indicate summation over all possible transitions
from the incoming to the outgoing wave.

\subsubsection{Particle / antiparticle transition, spin flip}

In the classical case the transition matrix elements for changing
from a particle into an antiparticle, and vice versa, are trivially
zero. Here we have additional factors $\phi$ and $\underline{\phi}$
to be tested. Using definitions for the $u_{k}^{\pm}$ we have:
\begin{align}
\left(u_{1}^{+}\right)^{\dagger}\gamma_{0}u_{2}^{+} & =\frac{m_{0}+E}{2m_{0}}\left(1,0,\frac{-p_{3}}{m_{0}+E},\frac{-p_{1}+ip_{2}}{m_{0}+E}\right)\gamma_{\phi^{2}}\gamma_{0}\left(\begin{array}{c}
-\phi p_{3}/\left(m_{0}+E\right)\\
\phi\left(-p_{1}-ip_{2}\right)/\left(m_{0}+E\right)\\
\underline{\phi}\\
0
\end{array}\right)=\phi\frac{-p_{3}+p_{3}}{2m_{0}}=0,\\
\left(u_{1}^{+}\right)^{\dagger}\gamma_{0}u_{2}^{-} & =\frac{m_{0}+E}{2m_{0}}\left(1,0,\frac{-p_{3}}{m_{0}+E},\frac{-p_{1}+ip_{2}}{m_{0}+E}\right)\gamma_{\phi^{2}}\gamma_{0}\left(\begin{array}{c}
\phi\left(-p_{1}+ip_{2}\right)/\left(m_{0}+E\right)\\
\phi p_{3}/\left(m_{0}+E\right)\\
0\\
\underline{\phi}
\end{array}\right)=\phi\frac{-p_{1}+ip_{2}+p_{1}-ip_{2}}{2m_{0}}=0.
\end{align}
The other six cases of the $u_{\mathrm{fin}}^{\dagger}\gamma_{0}u_{\mathrm{ini}}$
where $k_{\mathrm{fin}}\neq k_{\mathrm{ini}}$ are symmetric through
reordering of the vector components, as well as swapping factors in
the commutative product. A particle cannot change into an antiparticle
through elastic scattering, and vice versa.

Spin flip is also excluded:
\begin{align}
\left(u_{1}^{+}\right)^{\dagger}\gamma_{0}u_{1}^{+} & =\frac{m_{0}+E}{2m_{0}}\left(1,0,\frac{-p_{3}}{m_{0}+E},\frac{-p_{1}+ip_{2}}{m_{0}+E}\right)\gamma_{\phi^{2}}\gamma_{0}\left(\begin{array}{c}
0\\
1\\
\left(-p_{1}+ip_{2}\right)/\left(m_{0}+E\right)\\
p_{3}/\left(m_{0}+E\right)
\end{array}\right)=0,\\
\left(u_{2}^{+}\right)^{\dagger}\gamma_{0}u_{2}^{+} & =\frac{m_{0}+E}{2m_{0}}\left(\frac{-\phi p_{3}}{m_{0}+E},\frac{\phi\left(-p_{1}+ip_{2}\right)}{m_{0}+E},\underline{\phi},0\right)\gamma_{\phi^{2}}\gamma_{0}\left(\begin{array}{c}
\phi\left(-p_{1}+ip_{2}\right)/\left(m_{0}+E\right)\\
\phi p_{3}/\left(m_{0}+E\right)\\
0\\
\underline{\phi}
\end{array}\right)=0.
\end{align}

\subsubsection{Spin and particle type remains unchanged}

Interaction of the particle with the central $\widetilde{A}=\left(-\widetilde{Q}/\left|\vec{x}\right|,0,0,0\right)$
potential may only change the particle's momentum distribution across
different angles, but not its spin or type. The only spin contribution
therefore is:
\begin{equation}
\frac{d\widetilde{\sigma}_{\mathrm{spin}}}{d\Omega}=\left|\left(u_{k,\mathrm{fin}}^{\pm}\right)^{\dagger}\gamma_{0}u_{k,\mathrm{ini}}^{\pm}\right|^{2}\textrm{ for any }k,\pm\textrm{ unchanged}.
\end{equation}

The calculation is symmetric for particles and antiparticles, for
spin up and spin down, as well as rotational in space around the axis
of incoming momentum respective to the target. It is therefore calculated
for a particle ($k=1$) with spin up ($+$) in the $\left(x_{1},x_{2}\right)$
plane ($p_{3,\mathrm{ini}}=p_{3,\mathrm{fin}}=0$), assuming energy
conservation ($E_{\mathrm{fin}}=E_{\mathrm{ini}}\equiv E$):
\begin{align}
\left(u_{1,\mathrm{fin}}^{+}\right)^{\dagger}\gamma_{0}u_{1,\mathrm{ini}}^{+} & =\frac{m_{0}+E}{2m_{0}}\left(1,0,0,\frac{-p_{1,\mathrm{fin}}+ip_{2,\mathrm{fin}}}{m_{0}+E}\right)\gamma_{\phi^{2}}\gamma_{0}\left(\begin{array}{c}
1\\
0\\
0\\
\left(-p_{1,\mathrm{ini}}-ip_{2,\mathrm{ini}}\right)/\left(m_{0}+E\right)
\end{array}\right)\\
 & =\frac{m_{0}+E}{2m_{0}}\left(1+\frac{\left(-p_{1,\mathrm{fin}}+ip_{2,\mathrm{fin}}\right)\phi^{2}\left(p_{1,\mathrm{ini}}+ip_{2,\mathrm{ini}}\right)}{\left(m_{0}+E\right)^{2}}\right)\\
 & =\frac{\left(m_{0}+E\right)^{2}-\phi^{2}\left|\vec{p}\right|^{2}\left(\cos\theta-i\sin\theta\right)}{2m_{0}\left(m_{0}+E\right)}.
\end{align}
The last line expressed the difference between initial and final momentum
in terms of scattering angle $\theta$, while taking advantage of
$p_{3}=0$:
\begin{align}
p_{1,\mathrm{fin}}p_{1,\mathrm{ini}}+p_{2,\mathrm{fin}}p_{2,\mathrm{ini}} & \equiv\vec{p}_{\mathrm{fin}}\vec{p}_{\mathrm{ini}}=\left|\vec{p}\right|^{2}\cos\theta,\\
p_{1,\mathrm{fin}}p_{2,\mathrm{ini}}-p_{2,\mathrm{fin}}p_{1,\mathrm{ini}} & \equiv\vec{p}_{\mathrm{fin}}\times\vec{p}_{\mathrm{ini}}=-\left|\vec{p}\right|^{2}\sin\theta.
\end{align}

Using the identities
\begin{align}
\cos^{2}\theta & =1-\sin^{2}\theta,\\
\cos\theta & =1-2\sin^{2}\frac{\theta}{2},
\end{align}
there is:
\begin{align}
\frac{d\widetilde{\sigma}_{\mathrm{spin}}}{d\Omega} & =\left|\left(u_{1,\mathrm{fin}}^{+}\right)^{\dagger}\gamma_{0}u_{1,\mathrm{ini}}^{+}\right|^{2}\\
 & =\left|\frac{\left(m_{0}+E\right)^{2}-\phi^{2}\left|\vec{p}\right|^{2}\left(\cos\theta-i\sin\theta\right)}{2m_{0}\left(m_{0}+E\right)}\right|^{2}\\
 & =\frac{\left(\left(m_{0}+E\right)^{2}-\phi^{2}\left|\vec{p}\right|^{2}\cos\theta\right)^{2}+\left(\phi^{2}\left|\vec{p}\right|^{2}\sin\theta\right)^{2}}{\left(2m_{0}\left(m_{0}+E\right)\right)^{2}}\\
 & =\frac{\left(m_{0}+E\right)^{4}-2\left(m_{0}+E\right)^{2}\phi^{2}\left|\vec{p}\right|^{2}\cos\theta+\phi^{4}\left|\vec{p}\right|^{4}\cos^{2}\theta+\phi^{4}\left|\vec{p}\right|^{4}\sin^{2}\theta}{\left(2m_{0}\left(m_{0}+E\right)\right)^{2}}\\
\textrm{...using } & \cos^{2}\theta=1-\sin^{2}\theta\textrm{ ...}\nonumber \\
 & =\frac{\left(m_{0}+E\right)^{4}-2\left(m_{0}+E\right)^{2}\phi^{2}\left|\vec{p}\right|^{2}\cos\theta+\phi^{4}\left|\vec{p}\right|^{4}}{\left(2m_{0}\left(m_{0}+E\right)\right)^{2}}\\
\textrm{...using } & \cos\theta=1-2\sin^{2}\frac{\theta}{2}\textrm{ ...}\nonumber \\
 & =\frac{\left(m_{0}+E\right)^{4}-2\left(m_{0}+E\right)^{2}\phi^{2}\left|\vec{p}\right|^{2}+\phi^{4}\left|\vec{p}\right|^{4}+4\left(m_{0}+E\right)^{2}\phi^{2}\left|\vec{p}\right|^{2}\sin^{2}\frac{\theta}{2}}{\left(2m_{0}\left(m_{0}+E\right)\right)^{2}}\\
 & =\frac{\left(\left(m_{0}+E\right)^{2}-\phi^{2}\left|\vec{p}\right|^{2}\right)^{2}}{\left(2m_{0}\left(m_{0}+E\right)\right)^{2}}+\frac{\phi^{2}\left|\vec{p}\right|^{2}\sin^{2}\frac{\theta}{2}}{m_{0}^{2}}\\
 & =\frac{\left(m_{0}^{2}+2m_{0}E+E^{2}-\phi^{2}\left|\vec{p}\right|^{2}\right)^{2}}{\left(2m_{0}\left(m_{0}+E\right)\right)^{2}}+\frac{\phi^{2}\left|\vec{p}\right|^{2}\sin^{2}\frac{\theta}{2}}{m_{0}^{2}}\\
\textrm{...using } & m_{0}^{2}=E^{2}+\phi^{2}\left|\vec{p}\right|^{2}\textrm{ ...}\nonumber \\
 & =\frac{\left(2E\left(m_{0}+E\right)\right)^{2}}{\left(2m_{0}\left(m_{0}+E\right)\right)^{2}}+\frac{\phi^{2}\left|\vec{p}\right|^{2}\sin^{2}\frac{\theta}{2}}{m_{0}^{2}}\\
 & =\frac{E^{2}}{m_{0}^{2}}+\phi^{2}\frac{\left|\vec{p}\right|^{2}\sin^{2}\frac{\theta}{2}}{m_{0}^{2}}\\
\textrm{...using } & \frac{E^{2}}{m_{0}^{2}}=\frac{1}{1+\phi^{2}\left|\vec{v}\right|^{2}},\qquad\frac{\left|\vec{p}\right|^{2}}{E^{2}}=\left|\vec{v}\right|^{2}\textrm{ ...}\nonumber \\
 & =\frac{1+\phi^{2}\left|\vec{v}\right|^{2}\sin^{2}\frac{\theta}{2}}{1+\phi^{2}\left|\vec{v}\right|^{2}}.
\end{align}
This correctly recovers the classical case for $\left.\phi^{2}\right|_{\mathrm{QED}}=-1$.

\subsection{Result and comparison with the classical case}

Putting the (semi-)classical and spin contributions to the scattering
cross section together, the result is:
\begin{align}
\frac{d\widetilde{\sigma}}{d\Omega} & =\frac{d\widetilde{\sigma}_{\mathrm{class}}}{d\Omega}\frac{d\widetilde{\sigma}_{\mathrm{spin}}}{d\Omega}\\
 & \qquad\textrm{with }\frac{d\widetilde{\sigma}_{\mathrm{class}}}{d\Omega}=\frac{\left|\widetilde{q}\widetilde{Q}m_{0}\right|^{2}}{4\left|\vec{p}\right|^{4}\sin^{4}\frac{\theta}{2}},\qquad\frac{d\widetilde{\sigma}_{\mathrm{spin}}}{d\Omega}=\frac{1}{m_{0}^{2}}\left(E^{2}+\phi^{2}\left|\vec{p}\right|^{2}\sin^{2}\frac{\theta}{2}\right)=\frac{1+\phi^{2}\left|\vec{v}\right|^{2}\sin^{2}\frac{\theta}{2}}{1+\phi^{2}\left|\vec{v}\right|^{2}},\\
\Longrightarrow\qquad\frac{d\widetilde{\sigma}}{d\Omega} & =\frac{\left|\widetilde{q}\widetilde{Q}\right|^{2}}{4\left|\vec{p}\right|^{4}\sin^{4}\frac{\theta}{2}}\left(E^{2}+\phi^{2}\left|\vec{p}\right|^{2}\sin^{2}\frac{\theta}{2}\right)=\frac{\left|\widetilde{q}\widetilde{Q}m_{0}\right|^{2}}{4\left|\vec{p}\right|^{4}\sin^{4}\frac{\theta}{2}}\cdot\frac{1+\phi^{2}\left|\vec{v}\right|^{2}\sin^{2}\frac{\theta}{2}}{1+\phi^{2}\left|\vec{v}\right|^{2}},\\
\frac{d\sigma}{d\Omega} & =\left\Vert \frac{d\widetilde{\sigma}}{d\Omega}\right\Vert \equiv\left|\frac{d\widetilde{\sigma}}{d\Omega}\right|_{0}.\label{eq:rutherfordResult}
\end{align}

As compared to the classical result,
\begin{align}
\left.\frac{d\sigma}{d\Omega}\right|_{\mathrm{QED}} & =\frac{\left|qQ\right|^{2}}{4\left|\vec{p}\right|^{4}\sin^{4}\frac{\theta}{2}}\left(E^{2}-\left|\vec{p}\right|^{2}\sin^{2}\frac{\theta}{2}\right)\qquad\left(\textrm{for general }E,\left|\vec{p}\right|\right)\\
 & =\frac{\left|qQm\right|^{2}}{4\left|\vec{p}\right|^{4}\sin^{4}\frac{\theta}{2}}\cdot\frac{1-\left|\vec{v}\right|^{2}\sin^{2}\frac{\theta}{2}}{1-\left|\vec{v}\right|^{2}}\qquad\left(\textrm{only for }m>0,\left|\vec{v}\right|<1\right),
\end{align}
the following differences exist:
\begin{itemize}
\item Particle charge $\widetilde{q}$, target charge $\widetilde{Q}$,
and invariant mass $m_{0}$ may be rotated against one another in
$\mathbb{C}_{0}$ by arbitrary real angles. In the classical case,
charges may only have an opposite sign. For the purpose of this paper
this generalization to $\mathbb{C}_{0}$ has no predictive value,
since only the real absolute values of products $\widetilde{q}\widetilde{Q}$
or $\widetilde{q}\widetilde{Q}m_{0}$ appear in the $d\sigma/d\Omega$
result, making it impossible to separate the individual factors into
their respective magnitudes and phases in $\mathbb{C}_{0}$.
\item The phase $\phi^{2}$ appears in the spin-$\frac{1}{2}$ contribution
part $\phi^{2}\left|\vec{p}\right|^{2}\sin^{2}\frac{\theta}{2}$.
This makes for a different value in the predicted measurement outcome.
\end{itemize}
The ratio $r_{\alpha/\mathrm{QED}}$ between general (any $\alpha$)
and classical ($\alpha=\pi/2$) cross section for fixed charges $q$
and $Q$ is:
\begin{align}
r_{\alpha/\mathrm{QED}}:=\frac{d\sigma}{d\Omega}/\left.\frac{d\sigma}{d\Omega}\right|_{\mathrm{QED}} & =\frac{\left\Vert E^{2}+\phi^{2}\left|\vec{p}\right|^{2}\sin^{2}\frac{\theta}{2}\right\Vert }{E^{2}-\left|\vec{p}\right|^{2}\sin^{2}\frac{\theta}{2}}=\frac{\sqrt{E^{4}+2\cos\left(2\alpha\right)E^{2}\left|\vec{p}\right|^{2}\sin^{2}\frac{\theta}{2}+\left|\vec{p}\right|^{4}\sin^{4}\frac{\theta}{2}}}{E^{2}-\left|\vec{p}\right|^{2}\sin^{2}\frac{\theta}{2}}.
\end{align}
On first look, $r_{\alpha/\mathrm{QED}}$ is larger the higher the
particle's incident speed is ($E^{2}\approx\left|\vec{p}\right|^{2}$,
$\left|\vec{v}\right|^{2}\rightarrow1$) and the closer the scattering
angle is backwards ($\theta\rightarrow\pi$).

\subsection{Approximation for slow moving incident particle}

For slow moving particles, $E\gg\left|\vec{p}\right|$, there is $\left|\vec{v}\right|=\left|\vec{p}\right|/E\ll1$
and therefore
\begin{equation}
\left.r_{\alpha/\mathrm{QED}}\right|_{\left|\vec{v}\right|\ll1}\approx1+\left[2\left|\vec{v}\right|^{2}\cos^{2}\alpha\right]\sin^{2}\frac{\theta}{2}+\ldots
\end{equation}
At nonclassical $\alpha$ there is $\cos^{2}\alpha>0$ and therefore
backscattering ($\theta\approx\pi$) is enhanced by a term proportional
to $\left|\vec{v}\right|^{2}$ and $\cos^{2}\alpha$. This behavior
can be compared, at least qualitatively, with scattering of a classical
body in Schwarzschild geometry. This has been investigated \cite{refBlackHoleCrossSection}
and confirms the qualitative agreement (section 4 figure 2 shows enhanced
nonvanishing backscattering).

\subsection{Approximation for fast incident particle}

The faster the incoming particle gets, $\left|\vec{v}\right|^{2}\rightarrow1$,
the smaller the difference between $\left|\vec{p}\right|$ and $E$
becomes. The ratio $r_{\alpha/\mathrm{QED}}$ grows approximately
as
\begin{equation}
\left.r_{\alpha/\mathrm{QED}}\right|_{\left|\vec{v}\right|\rightarrow1}\approx\left[\frac{\cos^{2}\alpha}{1-\left|\vec{v}\right|}\right]\sin^{2}\frac{\theta}{2}.
\end{equation}
As $\left|\vec{v}\right|$ approaches the speed of light, scattering
becomes infinitely much stronger for general $\alpha$ relative to
the conventional case where $\cos^{2}\alpha=0$, growing asymptotically
like $1/x$ (with $x=1-\left|\vec{v}\right|$). This might become
a window to observability from quantum gravitational effects, as all
current assumptions for particle scattering assume a negligible contribution
from quantum gravity. The effect can be compared qualitatively with
the spacial distribution of hot inter- and intragalactic gas, and
estimated for lateral momentum transfer of a high-energy neutrino
through matter.

\subsection{Approximation for purely gravitational interaction}

Purely gravitational Rutherford scattering of a spin-$\frac{1}{2}$
particle may be estimated with all phases in $\mathbb{C}_{0}$ zero
(including $\alpha=0$), target charge to be its mass at rest $\widetilde{Q}:=M$,
and incident particle's charge its total energy $\widetilde{q}:=E$,
so that equation (\ref{eq:rutherfordResult}) becomes:
\begin{align}
\left.\frac{d\sigma}{d\Omega}\right|_{\mathrm{grav}} & =\frac{M^{2}E^{2}}{4\left|\vec{p}\right|^{4}\sin^{4}\frac{\theta}{2}}\left(E^{2}+\left|\vec{p}\right|^{2}\sin^{2}\frac{\theta}{2}\right)=\frac{M^{2}}{4\left|\vec{v}\right|^{4}\sin^{4}\frac{\theta}{2}}\cdot\left(1+\left|\vec{v}\right|^{2}\sin^{2}\frac{\theta}{2}\right).
\end{align}
As expected from the classical case, the purely gravitational elastic
scattering trajectory of the particle only depends on target charge
$M$ and particle speed $\vec{v}$, but not the particle's gravitational
charge\footnote{When particle and target are of comparable energy, e.g.~$1\,\mathrm{GeV}$
lepton on proton, target recoil would have to be taken into account.}. For fast moving particles $\left|\vec{v}\right|\approx1$ this becomes:
\begin{align}
\left.\frac{d\sigma}{d\Omega}\right|_{\mathrm{grav},\left|\vec{v}\right|\approx1} & \approx\frac{M^{2}}{4\sin^{4}\frac{\theta}{2}}\cdot\left(1+\sin^{2}\frac{\theta}{2}\right)=\frac{M^{2}}{4}\left(\frac{1}{\sin^{4}\frac{\theta}{2}}+\frac{1}{\sin^{2}\frac{\theta}{2}}\right).\label{eq:scatteringPurelyGravSpeedOfLight}
\end{align}

In order to compare this effect with the strength of other forces,
equation (\ref{eq:scatteringPurelyGravSpeedOfLight}) has to be multiplied
with $G^{2}/c^{4}\approx5.5*10^{-55}\,\mathrm{m}^{2}/\mathrm{kg}^{2}$
to obtain a magnitude in SI units. For (near) pointlike targets such
as electrons ($M_{e^{-}}\approx9.1*10^{-31}\,\mathrm{kg}$) or neutrons
($M_{n}\approx1.7*10^{-27}\,\mathrm{kg}$) this effect has a characteristic
length scale $GM/c^{2}$ in the order of $\sim10^{-53\ldots-56}\,\mathrm{m}$.
It will therefore be overshadowed by electromagnetic interaction and
inelastic scattering, where possible. Observable effects can only
be expected in the absence of these interactions (e.g.~elastic scattering
of neutrinos on a point mass) or at very large scales (e.g.~elastic
scattering of particles with gas in and around galaxies).

\subsection{Estimate for momentum transfer of a fast fermion}

This section estimates the momentum transfer during elastic scattering
of a fast incoming point-like fermion on a stationary fermion target.
Target recoil is neglected for simplification, as it will not change
the order of magnitude of the momentum transfer.

The radially symmetric cross section $\sigma_{m}$ is the area within
which an incident particle will be scattered at a minimum outgoing
angle $\theta_{m}$. Writing $d\Omega=2\pi\sin\theta\,d\theta$, equation
(\ref{eq:scatteringPurelyGravSpeedOfLight}) can be integrated:
\begin{eqnarray}
\sigma_{m}=\intop_{0}^{\sigma_{m}}d\sigma & = & \intop_{\theta_{m}}^{\pi}\frac{M^{2}}{4\sin^{4}\frac{\theta}{2}}\cdot\left(1+\sin^{2}\frac{\theta}{2}\right)2\pi\sin\theta\,d\theta.
\end{eqnarray}
Using
\begin{eqnarray}
\int\frac{\sin\theta}{\sin^{4}\frac{\theta}{2}}d\theta & = & -\frac{2}{\sin^{2}\frac{\theta}{2}}+\mathrm{const,}\\
\int\frac{\sin\theta}{\sin^{2}\frac{\theta}{2}}d\theta=2\int\cot\frac{\theta}{2}\,d\theta & = & 4\ln\left(\sin\frac{\theta}{2}\right)+\mathrm{const,}
\end{eqnarray}
this becomes:
\begin{eqnarray}
\sigma_{m} & = & \frac{\pi M^{2}}{2}\left[-\frac{2}{\sin^{2}\frac{\theta}{2}}+4\ln\left(\sin\frac{\theta}{2}\right)\right]_{\theta_{m}}^{\pi}\\
 & = & \frac{\pi M^{2}}{2}\left[\left(-2+0\right)-\left(-\frac{2}{\sin^{2}\frac{\theta_{m}}{2}}+4\ln\left(\sin\frac{\theta_{m}}{2}\right)\right)\right]\\
 & = & \pi M^{2}\left[\frac{1}{\sin^{2}\frac{\theta_{m}}{2}}-\left(1+\ln\left(\sin^{2}\frac{\theta_{m}}{2}\right)\right)\right].
\end{eqnarray}
Writing $\Delta p_{\theta}$ for momentum transfer at a given angle,
and relative momentum transfer $\Delta\hat{p}_{\theta}$ as
\begin{eqnarray}
\Delta p_{\theta} & = & 2\left|\vec{p}\right|\sin\frac{\theta}{2},\\
\Delta\hat{p}_{\theta} & := & \frac{\Delta p_{\theta}}{2\left|\vec{p}\right|}=\sin\frac{\theta}{2},
\end{eqnarray}
there is:
\begin{eqnarray}
\sigma_{m} & = & \pi M^{2}\left[\frac{1}{\left(\Delta\hat{p}_{\theta}\right)^{2}}-\left(1+\ln\left(\Delta\hat{p}_{\theta}\right)^{2}\right)\right],\\
\hat{\sigma}_{m}:=\frac{\sigma_{m}}{\pi M^{2}} & = & \frac{1}{\left(\Delta\hat{p}_{\theta}\right)^{2}}-\left(1+\ln\left(\Delta\hat{p}_{\theta}\right)^{2}\right).
\end{eqnarray}
Here, $\hat{\sigma}_{m}$ is the cross section relative to $\pi M^{2}$.
This can be solved for $\Delta\hat{p}_{\theta}$ using the Lambert
function (product log function) $W$ to:
\begin{eqnarray}
\left(\Delta\hat{p}_{\theta}\right)^{2} & = & \frac{1}{W\left(e^{\hat{\sigma}_{m}+1}\right)}.
\end{eqnarray}

The relative cross section $\hat{\sigma}_{m}$ is typically very large.
For example, assuming $\sigma_{m}$ the face area of a neutron with
radius $r_{n}\approx0.8*10^{-15}\mathrm{m}$, mass $M_{n}\approx1.7*10^{-27}\,\mathrm{kg}$,
and $G^{2}/c^{4}\approx5.5*10^{-55}\,\mathrm{m}^{2}/\mathrm{kg}^{2}$
there is approximately:
\begin{eqnarray}
\left.\hat{\sigma}_{m}\right|_{n} & \approx & \frac{\pi r_{n}^{2}}{\pi\frac{G^{2}}{c^{4}}M_{n}^{2}}\approx\frac{\left(0.8*10^{-15}\mathrm{m}\right)^{2}}{5.5*10^{-55}\,\mathrm{m}^{2}*\left(1.7*10^{-27}\right)^{2}}\approx\frac{0.64*10^{-30}}{5.5*10^{-55}*2.9*10^{-54}}\approx4*10^{77}.
\end{eqnarray}

For large argument, the Lambert function is roughly a logarithm, and
the relative momentum transfer $\Delta\hat{p}_{\theta}$ can be estimated:
\begin{eqnarray}
\lim_{x\rightarrow\infty}\frac{W\left(x\right)}{\ln x} & = & 1,\qquad\Longrightarrow\Delta\hat{p}_{\theta}\approx\frac{1}{\sqrt{\ln\left(e^{\hat{\sigma}_{m}+1}\right)}}\approx\frac{1}{\sqrt{\hat{\sigma}_{m}}}.
\end{eqnarray}

The impact parameter $b_{m}$ is defined to be at the radius of $\sigma_{m}$,
i.e.~the maximum distance from the direct path through the target:
\begin{eqnarray}
\sigma_{m} & = & \pi b_{m}^{2},\qquad\hat{\sigma}_{m}=\frac{\sigma_{m}}{\pi M^{2}}=\frac{b_{m}^{2}}{M^{2}}.
\end{eqnarray}
This allows to express the approximate relative momentum transfer
$\Delta\hat{p}_{m}$ in terms of a given impact parameter $b_{m}$
and target mass $A$ as:
\begin{eqnarray}
\Delta\hat{p}_{m} & \approx & \frac{M}{b_{m}}.
\end{eqnarray}
In SI units, this is:
\begin{eqnarray}
\Delta\hat{p}_{m} & \approx & \frac{GM}{c^{2}b_{m}}\approx7.4*10^{-27}\frac{\mathrm{m}}{\mathrm{kg}}*\frac{M}{b_{m}}.
\end{eqnarray}
As expected, the value is still very small compared to known particle
interactions through the other forces. When the number of interaction
partners becomes in the order of $10^{20}$ or more, or distances
are in the order of the diameter of the Milky Way ($\sim10^{20}\,\mathrm{m}$)
or larger, an effect may become observable. Given the number of uncertainties
on the exact workings of such scattering in the real world, the effect
here is small enough to not have been noticed yet.

\subsection{Qualitative prediction for intergalactic medium distribution}

Another opportunity for observing effects predicted in this work is
over large distance scales. While individual gravitational scattering
between particles is very weak, the difference as compared to current
model assumptions may become apparent over galactic distance scales.
Assuming that the origin of most highly energized particles and gas
in the universe is from within galaxies, increased backscattering
at higher energies as predicted in the calculations here should -
qualitatively - lead to a distribution of intergalactic gas that is
hotter and denser near galaxies as compared to theoretical model predictions.
Direct measurement is difficult, as such gas typically does not radiate
by itself (keywords include e.g.~``warm-hot intergalactic medium'',
``intracluster medium'', ``circumgalactic enrichment''). In recent
years, observation of absorption lines in the spectrum of remote quasars
(the ``Lyman-$\alpha$ forest'') has become a powerful tool to constrain
and tune theoretical models of galactic development. It is envisioned
that within the next decade or two it should be possible to make quantitative
comparisons between these observations and model predictions.

\section*{Acknowledgments}

With many thanks to Michael J.~Duff, Nichol Furey, and John Huerta
for helpful discussion and advice, and Wolfram Alpha for help with
series approximations. A special thanks to the organizers of the 4th
Mile High Conference on Nonassociative Mathematics (2017), University
of Denver, CO, during which the seeds for this work were planted.

\end{document}